\documentclass[12pt]{article}
\pdfoutput=1
\usepackage{subfigure}
\usepackage{amssymb,amsmath}
\usepackage{graphicx}
\usepackage{color}
\usepackage[colorlinks=true
,urlcolor=blue
,citecolor=blue
,linkcolor=blue
,pagecolor=blue
,linktocpage=true
,pdfproducer=medialab
]{hyperref}
\usepackage[a4paper,width=15.2cm]{geometry}
\makeatletter \renewcommand{\@dotsep}{10000} \makeatother
\usepackage{appendix}

\newcommand{\beq}{\begin{equation}}
\newcommand{\eeq}{\end{equation}}
\newcommand{\bea}{\begin{eqnarray}}
\newcommand{\eea}{\end{eqnarray}}

\def\lrprtmu{\stackrel{\leftrightarrow}{\partial_\mu}}

\begin{document}

\begin{center}

 {\Large\bf  Understanding Lorentz violation with Rashba interaction
 } \vspace{1cm}

{\large   Muhammad Adeel Ajaib\footnote{ E-mail: adeel@udel.edu}}

{\baselineskip 20pt \it
University of Delaware, Newark, DE 19716, USA  } \vspace{.5cm}

\vspace{1.5cm}
\end{center}

\begin{abstract}

Rashba spin orbit interaction is a well studied effect in condensed matter physics and has important applications in spintronics. The Standard Model Extension (SME) includes a CPT-even term with the coefficient $H_{\mu \nu}$ which leads to the Rashba interaction term. From the limit available on the coefficient  $H_{\mu \nu}$ in the SME we derive a limit on the Rashba coupling constant for Lorentz violation. In condensed matter physics the Rashba term  is understood as resulting from an asymmetry in the confining potential at the interface of two different types of semiconductors. Based on this interpretation we suggest that a possible way of inducing the $H_{\mu \nu}$ term in the SME is with an asymmetry in the potential that confines us to 3 spatial dimensions.

\end{abstract}

\newpage

\section{Introduction}\label{intro}

There are several examples  where a mechanism employed in particle physics can be realized in a condensed matter system. Spontaneous symmetry breaking, which is the basis of the Higgs mechanism, is one such example, which can be realized in ferromagnetic systems when spins,  at low temperatures,  align in a particular direction and the rotational symmetry of the system is broken. 
In this article we present another example where a well known empirical effect in condensed matter can serve as a guide to better understand the origin of a term that can lead to testable Lorentz violation. 

Lorentz symmetry implies the invariance of the laws of Physics and constancy of the speed of light in all inertial frames. There is no evidence for its violation to date, but scenarios exist in theory that incorporate this possibility and provide a framework that can be used to test this to a even higher degree of accuracy \cite{Pavlopoulos:1967dm, Kostelecky:1988zi}. Furthermore, evidence of Lorentz violation can also provide evidence for an underlying theory that exists at higher energy scales. The Standard Model Extension (SME) \cite{Colladay:1998fq}  is such a framework that incorporates the effects of Lorentz violation in a manner consistent with our current understanding of the standard model of particle physics. The various terms in the SME can arise as Vacuum Expectation Values (VEVs) of tensor fields in an underlying theory. String theory is an example of such an underlying theory, but there can possibly be other scenarios which can give rise to these terms in the SME. Herein, we shall discuss another possible scenario that can give rise to one of the terms in the SME.

Spintronics is an emerging field in condensed matter physics which involves designing innovative electronic devices by  manipulating the spin of the electron. Rashba spin orbit interaction (RSOI) has widespread applications in spintronics \cite{rashba-rev}.  This interaction term results from the asymmetry of the confining potential at the junction of, for example, two different semiconductors. 
The current understanding of the RSOI can lead to interesting implications for the SME which we will discuss in this article.

The paper is organized as follows: We briefly review the SME in section \ref{sme}. In section 
\ref{rashba-sme} we show that the Rashba term can be obtained from the non-relativistic limit of the Dirac equation with the $H_{\mu \nu}$ term. We also calculate the spinors and the Klein-Gordon equation in this section and obtain the dispersion relation. We discuss the RSOI term in section \ref{rashba} and derive a limit on the Rashba coupling for Lorentz violation from the known limit on the coefficient $H_{\mu \nu}$. In section \ref{rashba-interpret} we discuss the implication of the current understanding of the Rashba interaction term for the SME. We conclude in section \ref{conclude}.

\section{Brief review of the SME}\label{sme}

The Standard Model is now understood as an effective quantum field theory which is the low-energy limit of an underlying theory that describes gravity in addition to other forces. 
Lorentz violating extensions of the SM can be one of the possible ways to search for low-energy signals of such an underlying theory. The minimal SME (mSME) includes all SU(3)$\times$SU(2)$\times$U(1)  gauge invariant and renormalizable terms that violate particle Lorentz invariance (for  reviews see \cite{Kostelecky:2010ux, Bluhm:2005uj} and references therein).
 Scenarios that motivate the SME include mechanisms that arise in string theory where Lorentz tensor fields develop vacuum expectation values as a result of spontaneous breaking of Lorentz symmetry. Note that the terms in the SME preserve invariance of observer Lorentz transformations which involve a change  in coordinates. The violation of Lorentz invariance in the SME appears in the context of particle Lorentz transformations which involve boosting particles while the background fields remain invariant.

The QED sector of the mSME Lagrangian for a single species of fermion is given by ($\hbar=c=1$),
\begin{eqnarray}
{\cal L} =  {i \over 2} \bar{\psi}\Gamma^\mu \lrprtmu \psi -
\bar{\psi}M \psi ,
\label{smelag}
\end{eqnarray}
where, $\mu=0,1,2,3$ and
\begin{eqnarray}
\Gamma^\nu & = & \gamma^\nu + c^{\mu\nu}\gamma_\mu + d^{\mu\nu}\gamma_5 \gamma_\mu
+ e^\nu + if^{\nu}\gamma_5 + \frac12 g^{\lambda \mu \nu}
\sigma_{\lambda \mu}, \label{Gamma} \\
M & = & m + a_\mu\gamma^\mu + b_\mu\gamma_5 \gamma^\mu + \frac12
H_{\mu\nu}\sigma^{\mu\nu}\label{Mass}.
\end{eqnarray}
The operators with coefficients $a_{\mu}$, $b_{\mu}$, $e_{\mu}$, $f_{\mu}$ and $g_{\lambda \mu \nu}$ are CPT-odd, whereas the coefficients $H_{\mu \nu}$, $c_{\mu \nu}$ and $d_{\mu \nu}$ are associated with CPT-even operators. These coefficients are constant background fields which couple to fermions and can lead to observable Lorentz violating effects. In addition, these coefficients in the mSME are not functions of space-time. These fields remain invariant under CPT transformations, so, CPT is broken by operators that are odd under this symmetry. All these coefficients are expected to be very small since no evidence for Lorentz violation has been observed yet (for recent limits on these coefficients see \cite{Kostelecky:2008ts}). Furthermore, these fields transform in their respective manners (e.g. as vectors, tensors, etc.) under observer Lorentz transformations so that the  Lagrangian is observer Lorentz invariant. Under particle Lorentz transformations only the fermion fields are boosted or rotated and the background fields given in equations (\ref{Gamma}) and (\ref{Mass}) remain invariant. Therefore, the terms in the mSME lead to particle Lorentz violation. 

In this article we shall focus on the $H_{\mu \nu}$ term specifically the $0i$ ($i=1,2,3$) components in order to attain the Rashba   interaction term.
We consider the following Dirac Lagrangian with an additional term of the mSME,
\begin{eqnarray}
{\cal L}=\bar{\psi}(\mathit{i}   \gamma ^{\mu } \partial _{\mu }-m ) \psi- \frac{1}{2} H_{\mu \nu} \overline{\psi} \sigma^{\mu \nu} \psi.
\label{dirac1}
\end{eqnarray}
The coefficient $H_{\mu \nu}$ is real (${\cal L}^\dagger={\cal L}$), has dimensions of mass, and is antisymmetric, with the operator $\overline{\psi} \sigma^{\mu \nu} \psi$  even under CPT (C-odd, P-even and T-odd). 
A possible way of inducing the $H_{\mu \nu}$ term is through the VEV of an antisymmetric tensor field \cite{Altschul:2009ae}. 
Following limit on this coefficient is known from experiments involving Torsion pendulums and Xe/He masers \cite{Kostelecky:2008ts},
\begin{equation}
H_{\mu \nu} \lesssim 10^{-26} \mathrm{\ GeV}.
\label{limit}
\end{equation}
From this limit, we will calculate the limit on the Rashba coupling constant for Lorentz violation in section \ref{rashba} and compare it with limits known on the Rashba coupling constants for different materials in condensed matter. 


\section{Rashba interaction term in the SME}\label{rashba-sme}

In this section we will consider a simple case with a single non-zero component of the coefficient $H_{\mu \nu}$ and calculate the non-relativistic limit, Klein Gordon equation and spinors. The non-relativistic limit of various terms in the mSME given in equations (\ref{Gamma}) and (\ref{Mass}) have been calculated in \cite{Kostelecky:1999zh, Lehnert:2004ri} using the Foldy-Wouthuysen transformation and for some terms in \cite{Ferreira:2006kg}.  We choose the coefficients $H_{ij}=0$. Note that non-zero $H_{ij}$ coefficients can lead to terms of the form $\epsilon_{jkl} H_{kl}\sigma^j$ in the non-relativistic limit which can effect the Zeeman or hyperfine transitions in hydrogen  atoms \cite{Bluhm:2000gv}. We shall focus on the case of non-zero $H_{0i}$ components and show that anyone or all of these can lead to the RSOI term. As a simple case, we choose $H_{01}=H_{02}=0$ and $H_{03}\equiv h \neq 0$, which  amounts to choosing an observer Lorentz frame in which $H_{03}$ is the only non-zero coefficient. With these choices the equation associated with the Lagrangian (\ref{dirac1}) becomes 
\begin{eqnarray}
(\mathit{i}   \gamma ^{\mu } \partial_{\mu }-m ) \psi - h  \sigma^{03} \psi=0.
\label{dirac2}
\end{eqnarray}
\textbf{Non-relativistic limit:} In the non-relativistic limit, equation (\ref{dirac2}) yields a term that is of the form of RSOI. In order to get the non-relativistic limit we assume plane wave dependence of the wave-function as, $\psi=e^{-i p.x} w(\vec p)$, $w(\vec p)$ being a four component spinor in this case. Equation (\ref{dirac2}) in momentum space becomes
\begin{eqnarray}
(   \gamma ^{\mu } p_{\mu }-m - h  \sigma^{03} ) w(\vec p)=0.
\label{dirac3}
\end{eqnarray}
We choose  $w(\vec p)$ in two component form as $w = (\phi \ \chi)^T$, where $\phi$ and $\chi$ are two component spinors. We further assume the presence of an electromagnetic field and replace $p_{\mu} \rightarrow p_{\mu}-e A_{\mu}$, $A_{\mu}$ being the vector potential and $e$  the charge of the electron. Substituting $w$ and $p_{\mu}$ in equation (\ref{dirac3}), we get the following equations for the components $\phi$ and $\chi$
\begin{eqnarray}
\chi=\frac{\vec \sigma.\vec \Pi - i h  \sigma_3 }{E+m  -e A^0} \phi, \\
\phi=\frac{\vec \sigma.\vec \Pi + i h  \sigma_3 }{E-m  -e A^0} \chi,
\label{nr2}
\end{eqnarray}
where, $\vec \Pi= \vec p -e \vec A$ and $\vec \sigma$ are the Pauli matrices. In the non-relativistic limit, $E+m \cong 2 m $ and $E-m \cong H$, where $H$ is the non-relativistic Hamiltonian. Using these approximations, we get the following Hamiltonian with the Rashba interaction term
\begin{eqnarray}
H \phi &=&  \left[\frac{(\vec \sigma.\vec \Pi)^2}{2m} +\frac{h}{m}(\sigma_1 \Pi_2-\sigma_2 \Pi_1)+\frac{h^2 }{2m}  +e A^0\right]\phi  \\
&=& \left[\frac{\vec \Pi^2}{2m}- \frac{e}{2m} \vec \sigma.\vec B  +e A^0+\frac{h}{m}(\sigma_1 \Pi_2-\sigma_2 \Pi_1)+\frac{h^2 }{2m} \right]\phi,
\label{dirac5}
\end{eqnarray}
where the last two terms result from Lorentz violation and the second last term is the RSOI term for confinement in the z-direction. For $B=0$ and $A^0=0$, the eigen values of the above Hamiltonian are ($1\equiv x, 2 \equiv y, 3 \equiv z$)
\begin{eqnarray}
E=\frac{\vec p^2}{2m}\pm \frac{h}{m}  \sqrt{p_x^2+p_y^2}+\frac{h^2 }{2m},
\label{Enr}
\end{eqnarray}
which shows that the degeneracy of the spin up and down state is lifted because of the Rashba interaction term. This splitting can be seen as due to the presence of an effective magnetic field.  We shall discuss the Rashba interaction term in more detail in the next section.
 
\textbf{Klein Gordon equation:} The Klein Gordon equation for (\ref{dirac2}) can be obtained by multiplying it with the opposite sign of the mass term. This gives
\begin{eqnarray}
(\partial^{\mu} \partial_{\mu}+i h \{\gamma^{\mu},\sigma^{03} \}\partial_{\mu}+(h^2+m^2))\psi(x)=0.
\end{eqnarray}
This equation contains off-diagonal elements, which can be removed if we perform another multiplication, but this time with the opposite sign of the off-diagonal term yielding:
\begin{eqnarray}
(\left[ \partial^{\mu} \partial_{\mu} + (h^2+m^2) \right]^2+4 h^2
 (\partial_1^2+\partial^2_2))\psi(x)=0.
\end{eqnarray}
Assuming plane wave solutions, the dispersion relation comes out as
\begin{eqnarray}
\left[ p^2 - (h^2+m^2) \right]^2-4 h^2 (p_1^2+p^2_2)=0.
\end{eqnarray}
The values of energy for the above dispersion relation are:
\begin{eqnarray}
& &{E^2}=\left(\sqrt{p_1^2+p^2_2} \pm h \right)^2 + p_3^2 +m^2,  \\
& &\Rightarrow E=\pm \sqrt{p'^2 +p_3^2 + m^2 },
\label{disp1}
\end{eqnarray}
where, $p'=\sqrt{p_1^2+p^2_2} \pm h$. The negative energy solutions  are interpreted in the usual way as positive energy anti-particles. From the dispersion relation we can see that the degeneracy of the spin up and down states is lifted for both positive and negative energy solutions. Equation (\ref{disp1}) can be the relativistic dispersion relation for an electron confined in a 2-dimensional plane. It shows that the net momentum in the xy-plane is changed where as the momentum along the z-direction, remains unchanged. The momentum in the z-direction will be quantized for confinement in a potential well and can be ignored for a 2-dimensional electron gas (2DEG). Similar dispersion relations can be obtained for confinement in other planes. We saw from equation (\ref{Enr}) that the non-relativistic limit  of the modified Dirac equation also leads to the splitting of the energies of the two spin states. So in this case the splitting also takes place for relativistic particles, as can be seen from the energy dispersion relation (\ref{disp1}).

\textbf{Spinors:} The spinors for the Dirac equation (\ref{dirac2}) are given in Appendix A and can be chosen by first assuming plane wave solutions for particles as $\psi=e^{-i p.x} u(\vec p)$ and then choosing the two positive energy eigenstates which are linearly independent. The spinors for antiparticles are similarly chosen by assuming plane wave solutions with opposite four momenta as, $\psi=e^{i p.x} v(\vec p)$ and then choosing the positive energy solutions following the interpretation of negative energy solutions as positive energy anti-particles.  The spinors $u_1(\vec p)$ and $v_1(\vec p)$ are related by the charge conjugation operation as $u^c_1(\vec p,h)= i \gamma_2 u^*_1(\vec p,h) \propto v_1(\vec p,-h)$. Note the change in sign of the $h$ parameter. This is because the SME operator we consider is odd under CPT, the antiparticles are interpreted as interacting with the opposite sign background field $H_{\mu \nu}$. Furthermore, the charge conjugated Dirac equation has the opposite sign for this term since this operator is odd under charge conjugation. 
The spinors corresponding to spin up and down particles are also connected via the combined time reversal and parity operation as $u_1^{PT}(\vec p,h) = - \gamma_1 \gamma_3 (\gamma_0 e^{i\phi}) u^*_1(\vec p,h) \propto u_2(\vec p,-h)$. Again, the opposite sign for the parameter is due to the term being odd under time reversal operation. 

\section{Rashba Interaction}\label{rashba}
 Spin-orbit interaction is a relativistic effect which can be obtained from the non-relativistic limit of the Dirac equation, if we consider $O(v^2/c^2)$ terms, usually referred to as the Pauli SO term. It is given by (not using natural units),
\begin{eqnarray}
H_{SO}= - \frac{ \hbar}{4 m_e^2 c^2} \vec{\sigma} . (\vec{p} \times \vec{\nabla} V),
\label{soc}
\end{eqnarray}
where $m_e$ is the mass of the electron, $\vec{p}$ is its momentum, $V$ is the potential and $\mathbf{\sigma}$ are the Pauli matrices. The SOI is enhanced in materials like semiconductors due to large gradients in potentials \cite{rashba-rev}. Rashba interaction is a type of spin-orbit interaction (SOI) which exists in materials with structural inversion asymmetry \cite{rashba-rev, Rashba:1960, Rashba:1984, Winkler}. This asymmetry can be present at the surface of a crystal or at the interface of two different types of crystals where the space inversion symmetry of the material is broken. The asymmetry of the confining potential mimics an electric field perpendicular to the interface of the two materials and can influence a 2DEG trapped at this interface. Electrons, upon Lorentz transformation to their rest frame, `see' this electric field as an effective in-plane magnetic field with which their spin interacts to give rise to this SOI. This, therefore results in a spin polarized 2DEG at the interface and lifts the degeneracy of the two spin states of the electrons thereby splitting the fermi level into two as given in equation (\ref{Enr}). This effect can be tuned by a gate voltage \cite{nitta, Studer, Grundler} and  can therefore be used to control the spin state of the 2DEG.
The RSOI Hamiltonian for a 2DEG confined to move in the xy-plane is given by
\begin{eqnarray}
H_R= \frac{ \alpha_R}{\hbar}(\sigma_x p_y -\sigma_y p_x),
\label{rashba1}
\end{eqnarray}
where $\alpha_R$ is the Rashba coupling constant and is a measure of the asymmetry of the confining potential. The above Hamiltonian results in different energies for the spin up and down states for a 2DEG as can be seen from equation (\ref{Enr}). An electron moving in the x direction, for example, is either polarized in the +y or -y direction which can be seen as due to an effective field $B_y$ proportional to momentum $p_x$ ($\alpha_R p_x \propto \mu_B B_y$, $\mu_B$ being the Bohr magneton). The Rashba coupling constant is usually given in units of eV-m and is related to the coefficient $h$ by the following relation:
\begin{equation}
\alpha_R = \frac{\hbar c}{m} h.
\end{equation}
So the limit on the coefficient $h$ given in equation (\ref{limit}) translates to the following limit on the Rashba coupling constant for Lorentz violation ($\equiv \alpha_{RLV}$) for the electron,
\begin{equation}
\alpha_{RLV} \lesssim  10^{-30} \mathrm{\ eV\text{-}m}.
\end{equation}
 The experimental value of $\alpha_R$ for different materials is of the order $10^{-11}-10^{-10} \mathrm{\ eV\text{-}m}$. The limit available on $H_{\mu \nu}$, therefore, translates to a very small limit on the Rashba coupling constant. For the case of condense matter systems this constant is a measure of the structural inversion asymmetry and in our case, gives a measure of violation of Lorentz symmetry. The Rashba coupling constant for Lorentz violation is proportional to the background field $H_{\mu \nu}$. The splitting of the Fermi surfaces in this case will be really small because of the small value of the Rashba coefficient for Lorentz violation.

As described in section \ref{sme}, adding the LV term to the Dirac equation couples the electron to the constant background field $H_{\mu \nu}$. The origin of Rashba interaction in the mSME is due to these fields that transform in different ways under observer and particle Lorentz transformations. The Rashba interaction term results from the coupling of the $0i$ component  of the operator $\bar{\psi} \sigma_{\mu \nu} \psi$ with $H_{\mu \nu}$. The operator $\bar{\psi} \sigma_{\mu \nu} \psi$ breaks time reversal symmetry but not parity. Under the time reversal operation the non-relativistic limit in equation (\ref{disp1}) implies $E_+(-p) = E_-(p)$ and under parity $E_+(-p) = E_+(p)$. Under the combined operation of time reversal and parity operations, $E_+(p) = E_-(p)$, which implies degeneracy for the spin up and down states in a crystal. 
The Rashba interaction preserves time reversal symmetry but breaks parity.
 At the interface of two crystals the potential breaks the space inversion symmetry and leads to the RSOI.
Note that in the SME the operator that leads to the Rashba interaction term breaks time reversal symmetry and preserves parity.


\section{Interpretation of Rashba Interaction term in the SME}\label{rashba-interpret}

In this section we discuss how the current understanding of the Rashba interaction term in condensed matter can be utilized to propose a possible origin of the $0i$ components of the $H_{\mu \nu}$ term in the mSME. The Rashba term given in the Hamiltonian (\ref{dirac5}) results from confinement in the z-direction. Similarly if we add the non-zero terms for the coefficients $H_{02}$ and $H_{03}$ we can get a Rashba term for confinement in the x and y direction. The Hamiltonian in this case is given by \cite{Kostelecky:1999zh},
\begin{eqnarray}
{\cal H}=  \frac{\vec p^2}{2m}
+  \frac{1 }{2m} \epsilon_{ijk} H_{0i} \sigma_j p_k  
 + \frac{ H_{0i}^2  }{m} .
 \label{dirac7}
\end{eqnarray}
which can be written as,
\begin{eqnarray}
{\cal H}&=&\frac{\vec p^2}{2m}
+\frac{H_{01}}{m} (\sigma_2 p_3-\sigma_3 p_2) 
+\frac{H_{03}}{m} (\sigma_1 p_2-\sigma_2 p_1) \nonumber \\
& &+\frac {H_{02} }{m} (\sigma_3 p_1-\sigma_1 p_3) + \frac{ (H^2_{01}+H^2_{02}+H^2_{03})  }{2m} 
 .
\label{dirac7}
\end{eqnarray}
The energy eigenvalues for this case are,
\begin{eqnarray}
E=\frac{p^2}{2 m} \pm E'+ \frac{ H_{0i}^2  }{m}, 
\end{eqnarray}
where,
\begin{eqnarray}
E'= \frac{1}{m} \hspace*{-5mm} & &[(H_{02}  p_x-H_{01} p_y)^2+H_{03}^2 \left(p_x^2+p_y^2 \right)\\  \nonumber
& & -2 H_{03}  (H_{01}  p_x+H_{02}  p_y) p_z+\left(H_{02}^2+H_{01}^2\right) p_z^2 ]^{1/2}. \nonumber
\end{eqnarray}
The momentum dependent splitting of the electrons in this case is $2 E'$. Assuming the $H_{0i}$ coefficients of the same order,  $H_{01}/m \simeq H_{02}/m \simeq H_{03}/m \equiv \alpha_{RLV}$, $E'$ simplifies to
\begin{eqnarray}
E'= \alpha_{RLV} \sqrt{(  p_x- p_y)^2+(  p_y- p_z)^2+(  p_x- p_z)^2}.
\end{eqnarray}
Following the same procedure as section \ref{rashba-sme}, we can get the dispersion relation for this case as
\begin{eqnarray}
{E^2}= \overrightarrow{p}^2 +m^2 \pm 2 \alpha_{RLV} \sqrt{(  p_x- p_y)^2+(  p_y- p_z)^2+(  p_x- p_z)^2}   + 3 \alpha^2_{RLV}.
\label{disp2}
\end{eqnarray}
The relativistic dispersion relation shows that the spin degeneracy is lifted and the splitting of the spin up and down states depend on the momentum in all three directions.  

Keeping in view the manner in which the Rashba interaction term is interpreted in condensed matter physics we can propose a similar interpretation for the $H$ term in the mSME. The above Hamiltonian can be seen as resulting from confinement of the fermions in 3 dimensional space in a potential well.  The field $H_{\mu \nu}$ can therefore result from the asymmetry in this potential well. Rubakov and Shaposhnikov in \cite{Rubakov:1983bb} proposed that scalars and fermions can be trapped in 3 dimensional space in a potential well that is narrow along the fourth extra spatial dimension. 
This interpretation of generating the $H_{0i}$ component is inspired more from the empirical aspects of the Rashba interaction term. A more accurate way of showing this would be to begin with a potential that leads to both confinement of the matter fields and the asymmetry of which leads to the $H_{0i}$ component of the field $H_{\mu \nu}$.

Since the Rashba effect is an intensively studied term, this proposition can be tested  through very sensitive experiments. As discussed earlier, the RSOI leads to a spin polarized 2DEG at the interface of two semiconductors and the Rashba coupling can be controlled by a gate voltage. This leads to spin polarized currents and can be used to achieve control over the electron spin.   Rashba interaction is known to lead to the Spin Hall Effect \cite{SinovaPRL04}.  Similar to the ordinary Hall effect which leads to accumulation of opposite charges at two sides of a conductor, the SHE leads to the accumulation of opposite spins due to a spin current transverse to an applied electric current. Similarly, there are several devices that function on the RSOI such as the Datta Das transistor \cite{datta}.
If the background fields in the SME exist than a very sensitive experiment such as this can be devised in order to test the presence of the field $H_{\mu \nu}$ and, possibly, get a more stringent limit on this coefficient.  The presence of a non-zero field $H_{\mu \nu}$ can, therefore, not only serve as evidence for Lorentz violation, but might also hint towards the presence of extra spatial dimensions.

\section{Conclusion}\label{conclude}
We discussed a term in the SME that leads to the Rashba spin orbit interaction. The Rashba effect has wide spread applications in spintronics and has its origin in the asymmetry of the confining potential of a 2 dimensional electron gas. The limit on the coefficient $H_{\mu \nu}$, which yields the Rasha term, implies a very small limit on the derived Rashba coupling constant for Lorentz violation. Since the origin of the Rashba interaction is the asymmetry in the confining potential well, we proposed that the origin of the $H$ field in the SME can analogously be seen as due to the asymmetry in the potential well that confines us to 3+1 dimensional space-time. 

\section{Acknowledgments}
The author would like to thank Alan Kostelecky for useful discussions and suggestions. The author also greatly appreciates Emmanuel Rashba's useful comments on the draft. Finally, I would also like to thank Branislav Nikolic, Farzad Mahfouzi and Fariha Nasir for useful discussions.

\section*{Appendix A} 
\setcounter{equation}{0}  

\renewcommand{\theequation}{A-\arabic{equation}}

The spinors for the Dirac equation (\ref{dirac2}) for particles and anti-particles can be chosen as
\begin{eqnarray}
u_1(\vec p)&=& N^u_1\left(
\begin{array}{cc}
1 \\
\frac{-i( p_x + i p_y)}{\sqrt{p_x^2+p_y^2}} \\
\frac{-i(E_+-m)}{h- i p_z+\sqrt{p_x^2+p_y^2}} \\
\frac{p_x +i p_y}{\sqrt{p_x^2+p_y^2}} \frac{(E_+ -m)}{h- i p_z+\sqrt{p_x^2+p_y^2}}\\
\end{array} \right), \ \  
u_2(\vec p)=N^u_2\left(
\begin{array}{cc}
\frac{-i (p_x -i p_y)}{\sqrt{p_x^2+p_y^2}} \\
1 \\
\frac{p_x -i p_y}{\sqrt{p_x^2+p_y^2}} \frac{E_- -m}{-h+ i p_z+\sqrt{p_x^2+p_y^2}}\\
\frac{-i(E_- -m)}{-h+ i p_z+\sqrt{p_x^2+p_y^2}} \\
\end{array} \right), \nonumber \\
v_1(\vec p)&=& N^v_1 \left(
\begin{array}{cc}
\frac{p_x -i p_y}{\sqrt{p_x^2+p_y^2}} \frac{E_- -m}{-h+ i p_z+\sqrt{p_x^2+p_y^2}}\\
\frac{-i(E_- -m)}{-h+ i p_z+\sqrt{p_x^2+p_y^2}} \\
\frac{-i( p_x - i p_y)}{\sqrt{p_x^2+p_y^2}} \\
1 \\
\end{array} \right), \ \  
v_2(\vec p)= N^v_2 \left(
\begin{array}{cc}
\frac{-i(E_+ -m)}{h- i p_z+\sqrt{p_x^2+p_y^2}} \\
\frac{p_x +i p_y}{\sqrt{p_x^2+p_y^2}} \frac{E_+ -m}{h- i p_z+\sqrt{p_x^2+p_y^2}}\\
1 \\
\frac{-i( p_x + i p_y)}{\sqrt{p_x^2+p_y^2}} \\
\end{array} \right). \nonumber
\label{spinors1} 
\end{eqnarray}
These can be written in a more compact form as
\begin{eqnarray}
u_1(\vec p)&=& N^u_1\left(
\begin{array}{cc}
\chi_1 \\
F \chi'_1 \\
\end{array} \right), \ \  
u_2(\vec p)=N^u_2 \left(
\begin{array}{cc}
 \chi_2 \\
- G \chi'_2 \\
\end{array} \right), \nonumber \\
 v_1(\vec p)&=&N^v_1 \left(
\begin{array}{cc}
-G  \chi'_2 \\
 \chi_2 \\
\end{array} \right), \ \  
v_2(\vec p)=N^v_2 \left(
\begin{array}{cc}
F \chi'_1 \\
 \chi_1 \\
\end{array} \right)
\label{spinors2} 
\end{eqnarray}
where $N_{1,2}^{u,v}$ are spinor normalization factors, $\chi'_i=\sigma_3 \chi_i$ and
\begin{eqnarray}
\chi_1(\vec p)&=& \left(
\begin{array}{cc}
1 \\
\frac{-i( p_x +i p_y)}{\sqrt{p_x^2+p_y^2}} \\
\end{array} \right), \ \  
\chi_2(\vec p)=\left(
\begin{array}{cc}
\frac{-i (p_x -i p_y)}{\sqrt{p_x^2+p_y^2}} \\
1 \\
\end{array} \right) \\
F &=& \frac{- i(E_{+}-m)}{ h- i p_z+\sqrt{p_x^2+p_y^2}},  \\
G &=& \frac{- i(E_{-}-m)}{ -h+ i p_z+\sqrt{p_x^2+p_y^2}}, \\
%
{E_{\pm}}&=&\sqrt{ (\sqrt{p_1^2+p^2_2} \pm h )^2 + p_3^2 +m^2 }. 
\label{disp1a}  
\end{eqnarray}
The spinors $u_1(\vec p)$ and $u_2(\vec p)$ correspond to particles, whereas $v_1(\vec p)$ and $v_2(\vec p)$ correspond to anti-particles. The spinors $u_1(\vec p)$ and $v_1(\vec p)$ are related by the charge conjugation operation as $u^c_1(\vec p,h)= i \gamma_2 u^*_1(\vec p,h) \propto v_1(\vec p,-h)$. The spinor corresponding to spin up particles $u_1(\vec p)$ is related to the spinor with opposite spin  $u_2(\vec p)$ by the time reversal and parity operator, i.e., $u_1^{PT}(\vec p,h) = - \gamma_1 \gamma_3 (\gamma_0 e^{i\phi}) u^*_1(\vec p,h) \propto u_2(\vec p,-h)$, where the proportionality constant involve the normalization constants and $\phi$ is an arbitrary phase. The orthonormality conditions for the spinors are given by
\begin{eqnarray}
u^{(1)\dagger}({\vec p})u^{(1)}({\vec p}) &=&  \frac{4(E_+-m)}{E_+} ,\nonumber \\
u^{(2)\dagger}({\vec p})u^{(2)}({\vec p}) &=& 2+\frac{2(E_--m)^2}{E_-^2} ,\nonumber \\
v^{(1)\dagger}({\vec p})v^{(1)}({\vec p}) &=& 2+\frac{2(E_--m)^2}{E_-^2} ,\nonumber \\
v^{(2)\dagger}({\vec p})v^{(2)}({\vec p}) &=& \frac{4(E_+ -m)}{E_+} ,\nonumber\nonumber \\
u^{(r)\dagger}({\vec p})u^{(r')}({\vec p}) &=& 0, (r \neq r')\nonumber \\
v^{(r)\dagger}({\vec p})v^{(r')}(-{\vec p}) &=& 0, (r \neq r') \nonumber \\
u^{(r)\dagger}({\vec p})u^{(r')}({\vec p}) &=& 0,\nonumber \\
v^{(r)\dagger}(-{\vec p})u^{(r')}({\vec p}) &=& 0 ,
\label{ortho-conds}
\end{eqnarray}
where $r,r'=1,2$.


\end{document}